\documentclass[runningheads]{svmult}

\usepackage{makeidx}   % allows index generation
\usepackage{graphicx}  % standard LaTeX graphics tool
\usepackage{subeqnar}  % subnumbers individual equations
\usepackage{multicol}  % used for the two-column index
\usepackage{physprbb}  % modified textarea for proceedings,
\makeindex             % used for the subject index
\def\lsim{\raise0.3ex\hbox{$<$}\kern-0.75em{\lower0.65ex\hbox{$\sim$}}}
\def\gsim{\raise0.3ex\hbox{$>$}\kern-0.75em{\lower0.65ex\hbox{$\sim$}}}

\begin{document}
\title*{Interaction Models for GRB Afterglows} 
\toctitle{Ambient Interaction Models for GRBs}
% allows explicit linebreak for the table of content
%
%
\titlerunning{GRB Afterglow Models} 
% allows abbreviation of title, if the full title is too long
% to fit in the running head
%
\author{Zhi-Yun Li
\and Roger A. Chevalier }
\authorrunning{Li and Chevalier} 
% if there are more than two authors,
% please abbreviate author list for running head
%
%
\institute{Department of Astronomy, University of Virginia, Charlottesville,
VA 22903, USA}

\maketitle              % typesets the title of the contribution

\begin{abstract}
We review   interaction models for GRB afterglows, with an eye on 
constraining the nature of their progenitors and the geometry of 
explosion. Evidence is presented for two types of progenitors. 
The radio afterglow of GRB 980425/SN 1998bw and the multi-frequency
observations of GRB 970508 can be fitted by a blast wave expanding
into a wind-type medium, pointing to a massive star progenitor. 
The broadband afterglow data of GRBs 990123 and 990510 are better 
modeled by a jet expanding into a constant-density medium, implying 
a compact star merger origin. 
Among other well observed GRBs, the jet model appears to be
more widely applicable than the wind model, although
some cases are ambiguous.
The model fits often require a deviation of the energy distribution 
of the radiating electrons from the commonly assumed single 
power-law form, particularly for sources with a rapid decline and/or 
pronounced steepening in the optical light curves. Transition to 
non-relativistic evolution has been suggested as an alternative 
explanation for the light curve steepening, although to produce 
steepening on the order of days or less would require very high 
ambient densities, which are generally difficult to reconcile with 
radio observations. Major open issues include the hydrodynamics of 
jet-ambient medium interaction, a self-consistent determination of 
the electron energy distribution, and the effects of pair production
on the early afterglows which are expected to be particularly large 
for the wind interaction model.
\end{abstract}

%\noindent
%{\bf keywords}: gamma-ray burst, progenitor, afterglow, shockwave, 
% synchrotron, ambient medium

\section{Introduction} 
The simplest model of the afterglows of gamma-ray bursts (GRBs) involves 
a spherical relativistic blast wave expanding into a constant-density, 
presumably interstellar, medium  \cite{K94,MR97}. 
The afterglows are emitted by nonthermal electrons accelerated at the 
shock front to an energy distribution usually assumed to be a power 
law above some cutoff determined by the shock
velocity. The predicted power-law decay of the afterglow 
emission with time was subsequently observed at X-ray \cite{Ce97} 
 and optical \cite{vPe97} wavelengths for GRB 
970228, giving basic confirmation to this now ``standard'' picture 
\cite{W97a,WRM97}. 

To date, nearly 50 GRB afterglows have been observed at more than one
frequency (for reviews of afterglow observations see Chapters by Piro, 
Pian \& Fruchter, and Weiler). A dozen or so of these are well observed 
at multiple frequencies to allow for detailed modeling. Some features 
of these afterglows turn out to be difficult to accommodate in the 
standard model. The most noticeable is the relatively steep 
decline of the optical light curves, sometimes preceded by flatter
evolution. 
The steep decline is usually attributed to a collimated or jet-like, 
rather than spherical, initial energy injection \cite{R97,SPH99}.
 Alternatively, it could be due to the
transition of blast wave evolution to the non-relativistic regime 
\cite{DL99}, a non-standard electron energy distribution \cite{LC01,P01},
 and/or expansion into a wind-type 
ambient medium \cite{CL00}. The question of a wind versus 
a constant-density surrounding medium is a crucial one for the 
progenitors of GRBs, since massive stars, one of the leading 
candidates for GRB progenitors \cite{W93,P98},
should be surrounded by a wind. In contrast, GRBs resulted from 
compact star mergers, the other leading candidate, are expected to 
be surrounded by the general interstellar medium (see Chapter by 
Waxman). 

The plan of the Chapter is as follows.  
In \S~\ref{ambient}, we discuss the likely environments of GRBs
relevant to afterglow evolution.
In \S~\ref{model}, we  outline
the basics of various ambient interaction models for GRB afterglows. 
These models are then applied to individual GRBs with relatively well-observed 
afterglows (\S~\ref{application}). We find that some of the 
sources are probably wind-interactors, while others are better modeled 
as interacting with a constant-density medium. This and other results 
are discussed in \S~\ref{discussion}. 

\section{The Ambient Medium}
\label{ambient}

Models for the GRB afterglows indicate that the emission comes
from a region  $\sim 10^{16}-10^{18}$ cm from the source of the
explosion.
The nature of the material in this region depends on the GRB progenitors,
which are presently not known.
There are two main possibilities.
In one model, the progenitors are massive stars at the end of
their lives \cite{W93,P98}.
In this case, the interaction is expected to be primarily with
the mass loss from the progenitor leading up to the explosion.
The other class of object is the merger of two compact objects
for which there are a number of possibilities \cite{FWH99}.
In this case, there may be some debris from the merger, but the
interaction is expected to be primarily with the ISM (interstellar
medium).
We consider these possibilities in turn.

If the progenitors are massive stars, there is an analogy to
the explosions of core collapse supernovae, for which there
is abundant evidence that they are interacting with the winds from
the progenitor stars (see Chapter by Chevalier \& Fransson).
In general, the interaction appears to be with the free wind
from the progenitor star, with density
$\rho_w=\dot M/(4\pi r^2 v_w)$, where $\dot M$ is the mass loss
rate from the star and $v_w$ is the wind velocity.
In most of the supernova cases, the radial range that is observed
is out to a few $10^{17}$ cm, so that the mass loss characteristics
have not substantially changed during the time that mass is supplied
to the wind and $\rho_w\propto r^{-2}$.
The density in the wind depends on the type of progenitor.
Red supergiant stars, which are thought to be the progenitors of
most Type II supernovae, have slow dense winds.
Wolf-Rayet stars, which are thought to be the progenitors of
 Type Ib and Ic supernovae, have faster, lower density winds.
SN 1987A is a special case in which the star was a red supergiant,
but became a blue, B3 supergiant  about $10^4$ years before
the explosion (see Chapter by McCray).
The result was a complex circumstellar medium, including a
dense ring at a radius of $6\times 10^{17}$ cm.

If GRBs do have massive star progenitors, there are a number of
arguments suggesting that Wolf-Rayet stars are the most likely
progenitors:  (1) SN 1998bw, the best
case of a SN -- GRB association (GRB 980425), was of Type Ic, with a
probable Wolf-Rayet progenitor.
(2)  The high energy of GRBs suggest that a moderately massive %, $> 3 ~M_\odot$,
black hole is involved, which, in turn, requires a massive, $\gsim
 20-25 ~M_\odot$,
progenitor \cite{EH98}.
These stars are likely to be Wolf-Rayet stars at the end of their 
lives \cite{GS96}.
(3)  The relativistic flow from a central object may be able to penetrate
a relatively compact Wolf-Rayet star, but probably cannot
penetrate an extended red supergiant star \cite{MWH01}.
(4)  If a rapidly rotating black hole is required for the GRB
explosion, a merger at the center of a massive star is one way
to obtain a rapidly rotating core.
The merger process can give rise to a Wolf-Rayet star.
None of these arguments is definitive, but they do point toward
the most plausible progenitors.

The circumstellar medium created by a $35~M_{\odot}$ initial mass star
that passes through a Wolf-Rayet phase and a prior red supergiant
phase has been simulated by
Garcia-Segura et al. \cite{GS96}.
They find that during the Wolf-Rayet phase, a broken shell is created
at a radius of a few pc or more.
This type of shell has been observed around a number of
Wolf-Rayet stars.
Inside of the shell is a region of shocked wind, with an approximately
constant density \cite{CI83}.
The size of the shocked region is determined by what is required to
decelerate the wind and is typically less than half of the shell radius.
In general, the wind ram pressure $p=\rho_w v_w^2$ determines the
size of the wind bubble that is created.
The pressure is $p/k = 3.6\times 10^9~\dot M_{-5} v_8 r_{17}^{-2}$
cm$^{-3}$ K,
where $\dot M_{-5}=\dot M/10^{-5}~M_\odot ~\rm yr^{-1}$,
$v_8=v_w/10^8$ cm s$^{-1}$, and
$r_{17}=r/10^{17}$ cm.
For comparison, the interstellar pressure in the solar neighborhood
is $\sim 3000$ cm$^{-3}$ K.
The region out to the wind termination shock can be described by
a density $\rho_w = 5\times 10^{11} A_* r^{-2}$ g cm$^{-3}$, where
$A_*=1 \dot M_{-5} v_8^{-1}$.
Considering the radial range relevant to GRB afterglows,
we expect this to apply to most cases, unless the wind is especially
weak or the surrounding pressure is extraordinarily high.
Ramirez-Ruiz et al. \cite{RR00} have followed the circumstellar mass
loss from Wolf-Rayet stars and found that the free wind does not extend
to large radii in their treatment.

For bursts that interact directly with the ISM,
the size scale of the afterglows suggests that the medium has
a roughly constant density, although the ISM is known to have
considerable density inhomogeneity on a range of scales.
The range of interstellar densities in a galaxy like our own goes from
$\sim 10^{-3}$ cm$^{-3}$ in the hot ISM to $\gsim 10^6$ cm$^{-3}$
in compact cores in molecular clouds.
The high densities are found in a very small volume fraction so that
a typical event is unlikely to occur in such a region.
The situation may change in a starburst region, where a significant
fraction of the gas may have a density $\gsim 10^3$ cm$^{-3}$
\cite{So01}.
The radio-bright, compact supernova remnants in starburst regions
may be interacting with this dense interstellar component \cite{CF01}.
The finding that GRB 010222 may have occurred within a very active
starburst \cite{Fre01} indicates that the interstellar 
environments of GRBs may have
unusual properties.

Although the interstellar and wind models are the two main types
of environments considered for afterglows, there is a different
scenario involving a massive star, motivated by the observation
of possible Fe lines in the X-ray spectra of afterglows
\cite{Pe99,Ye99,Pire00,Ane00}.
The observations appear to require a substantial mass of Fe at
$r\sim 10^{16}$ cm, so one suggestion is that a supernova
occurs before a GRB and that the ejecta have time to expand to
this radius \cite{Ve01}.
This requires that the GRB be delayed by months.
The supernova would expand into the progenitor wind, creating a
complex circumburst region in the inner part of the wind.

\section{Ambient Interaction Models}
\label{model}
\subsection{Standard ISM-Interaction Model}
We start with the simplest model of GRB afterglows involving synchrotron 
emission from a spherical relativistic blast wave propagating into 
a constant-density interstellar medium. The model was worked out in 
some detail by M\'esz\'aros \& Rees \cite{MR97} in advance of the afterglow 
detection. It has since been elaborated upon by others (e.g., 
\cite{W97a,W97b,WG99}),
and in particular \cite{SPN98}. 
 Here, we will follow the formalism of Sari et al. \cite{SPN98}, which 
has been widely used to interpret the spectra and light curves of 
GRB afterglows. This standard model serves as a benchmark against 
which other models will be compared.

The standard model assumes that the synchrotron-emitting electrons are 
accelerated at the shock front to a power-law distribution of Lorentz 
factor $N(\gamma_{\rm e})~d\gamma_{\rm e}\propto\gamma_{\rm e}^{-p}~d
\gamma_{\rm e}$ (with $p > 2$) above some minimum cutoff $\gamma_{\rm m}$,
which is determined by the shock velocity. 
This distribution is further modified by synchrotron cooling
in the downstream flow. Denote
by $\gamma_{\rm c}$ the Lorentz factor of an electron which cools
in a blast wave expansion time. In the ``fast cooling'' case with 
$\gamma_{\rm c} < \gamma_{\rm m}$, 
the electron distribution is given by $\gamma_{\rm e}^{-2}$ 
between $\gamma_{\rm c}$ and $\gamma_{\rm m}$ and $\gamma_{\rm e}
^{-p-1}$ above $\gamma_{\rm m}$. This electron distribution produces
a synchrotron spectrum of the flux $F_{\nu}\propto \nu^{-1/2}$ between 
$\nu_{\rm c}$ and $\nu_{\rm m}$, the characteristic frequencies
of the photons produced by electrons with $\gamma_{\rm c}$ and 
$\gamma_{\rm m}$ respectively, $F_{\nu}\propto 
\nu^{-p/2}$ above $\nu_{\rm m}$, and $F_{\nu}\propto \nu^{1/3}$ below 
$\nu_{\rm c}$. The spectrum is modified by synchrotron
self-absorption, which typically occurs at relatively low frequencies. 
Assuming that the self-absorption frequency $\nu_{\rm a}$ is 
less than the cooling frequency $\nu_{\rm c}$ and that the 
absorbing electrons that have cooled for different lengths of 
time are well mixed spatially \cite{GPS00}, we have 
$F_{\nu}\propto \nu^2$ below $\nu_{\rm a}$. In the opposite, ``slow 
cooling'' case with $\gamma_{\rm c} > \gamma_{\rm m}$, the electron 
distribution steepens to $\gamma_{\rm e}^{-p-1}$ above $\gamma_{\rm c}$. 
Assuming that the self-absorption frequency $\nu_{\rm a}$ is less than 
the characteristic frequency $\nu_{\rm m}$, we have the following 
spectrum: $F_{\nu}\propto \nu^2$ for $\nu < \nu_{\rm a}$, $F_{\nu}
\propto\nu^{1/3}$ for $\nu_{\rm a} < \nu < \nu_{\rm m}$, $F_{\nu}
\propto\nu^{-(p-1)/2}$ for $\nu_{\rm m} < \nu < \nu_{\rm c}$, and 
$F_{\nu}\propto \nu^{-p/2}$ for $\nu > \nu_{\rm c}$. These 
broken power-law expressions are often used to interpret the 
instantaneous spectra of GRB afterglows. They tend to agree
with the more accurately determined spectra to within a factor
of a few (e.g., \cite{GPS99}).

To obtain light curves, one must first determine the time evolution of 
the characteristic frequencies ($\nu_{\rm a}$, $\nu_{\rm m}$ and
$\nu_{\rm c}$) and the peak flux $F_{\nu,{\rm max}}$. The evolution
depends on the dynamics of the blast wave. For the spherical blast wave 
propagating into a constant-density medium envisioned in the standard 
model, the dynamics are described by the self-similar solutions of 
Blandford \& McKee \cite{BM76}: the blast wave Lorentz factor $\gamma$ varies 
with the distance $R$ from the explosion center as $\gamma\propto
R^{-3/2}$ in the adiabatic regime and $\gamma\propto R^{-3}$ in the
fully radiative regime. The Lorentz factor and distance are related to 
the detector-time $t$ as $t\propto R/\gamma^2$. Making the standard
assumption that a constant fraction $\epsilon_{\rm e}$ ($\epsilon_{\rm 
B}$) of the blast wave energy goes into the electrons (magnetic fields)
and making use of the jump conditions for relativistic shocks, one 
obtains the following scalings: $\nu_{\rm a}\propto t^0$, $\nu_{\rm m}
\propto t^{-3/2}$, $\nu_{\rm c}\propto t^{-1/2}$, and $F_{\nu,{\rm max}}
\propto t^0$ for the slow cooling case, which is always in the adiabatic
regime. In the fast cooling case, if the electron energy fraction 
$\epsilon_{\rm e}$ is close to unity, then the blast wave is fully 
radiative, and the scalings become: $\nu_{\rm a}\propto t^{-4/5}$, 
$\nu_{\rm c}\propto t^{-2/7}$, $\nu_{\rm m}\propto t^{-12/7}$, and 
$F_{\nu,{\rm max}}\propto t^{-3/7}$. If on the other hand $\epsilon
_{\rm e}\ll 1$, the evolution 
remains adiabatic, and the scalings are the same as in the slow
cooling case, except for the self-absorption frequency which now 
decreases with time as $\nu_{\rm a}\propto t^{-1/2}$. These scalings
are combined with the instantaneous spectra listed above to obtain 
light curves.

The light curve at a given observing frequency $\nu$ can be described 
by a broken power-law with breaks at various characteristic times. 
These include the times $t_{\rm a}$, $t_{\rm m}$ and $t_{\rm c}$ 
when the characteristic frequencies $\nu_{\rm a}$, $\nu_{\rm m}$ 
and $\nu_{\rm c}$ pass, respectively, the observing frequency $\nu$. 
Ignoring self-absorption, Sari et al. \cite{SPN98} obtained two types of 
light curves in two frequency regimes, separated by a critical 
frequency $\nu_0$ at which the characteristic times $t_{\rm m}$ and 
$t_{\rm c}$ become equal, i.e., $t_{\rm m}(\nu_0)=t_{\rm c}(\nu_0)\equiv 
t_0$. The critical time $t_0$ divides the early, fast cooling part
of the light curve from the later, slow cooling part. In the high 
frequency regime where $\nu > \nu_0$, the ordering of the characteristic
times is $t_{\rm c} < t_{\rm m} < t_0$, which yields a light curve 
of $F_\nu\propto
t^{1/6}$ for $t < t_{\rm c}$, $F_\nu\propto t^{-1/4}$ for $t_{\rm c}
< t < t_{\rm m}$, and $F_\nu\propto t^{(2-3p)/4}$ for $t > t_{\rm m}$
in the case of adiabatic evolution. In the opposite case of 
fully radiative evolution, which is possible only before the critical 
time $t_0$ 
when the electrons are fast cooling, one has
instead $F_\nu\propto t^{-1/3}$ for $t < t_{\rm c}$, $F_\nu\propto 
t^{-4/7}$ for $t_{\rm c}< t < t_{\rm m}$, and $F_\nu\propto 
t^{(2-6p)/7}$ for $t_{\rm m} < t < t_0$. In the low frequency regime 
where $\nu < \nu_0$, the ordering of the characteristic times becomes 
$t_0 < t_{\rm m} < t_{\rm c}$. The light curve is then $F_\nu\propto 
t^{1/6}$ for $t < t_0$, $F_\nu\propto t^{1/2}$ for $t_0 < t < t_{\rm m}$, 
$F_\nu\propto t^{3(1-p)/4}$ for $t_{\rm m} < t < t_{\rm c}$, and 
$F_\nu\propto t^{(2-3p)/4}$ for $t > t_{\rm c}$ in the adiabatic case. 
In the fully radiative case, one has $F_\nu\propto t^{-1/3}$ for 
$t < t_0$ instead. 

Synchrotron self-absorption modifies the light curves.  We will 
concentrate on the case of adiabatic evolution here and below; it 
is more likely than the fully radiative case during most of the 
afterglow phase of a GRB evolution (e.g., \cite{MR97}). 
The self-absorption is characterized by the absorption frequency 
$\nu_{\rm a}$, which decreases with time as $\nu_{\rm a}
\propto t^{-1/2}$ during the fast cooling period $t < t_0$. 
It has a constant value, denoted by $\nu_{\rm a,0}$, between $t_0 < t < 
t_{\rm am}$, where $t_{\rm am}$ is another critical time
when the characteristic frequencies $\nu_{\rm a}$ and $\nu_{\rm m}$
become equal. In the frequency regime $\nu > \nu_{\rm a,0}$, both 
the high and low frequency light curves discussed above have an 
additional power-law segment $F_\nu\propto 
t$ before the characteristic time $t_{\rm a}$, which is smaller 
than any other characteristic time. Collecting all time and frequency 
dependences, we finally have 
$F_\nu\propto t \nu^2$ ($t < t_{\rm a}$), $F_\nu\propto t^{1/6}
\nu^{1/3}$ ($t_{\rm a} < t < t_{\rm c}$), $F_\nu\propto t^{-1/4}
\nu^{-1/2}$ ($t_{\rm c} < t < t_{\rm m}$), and $F_\nu\propto t^{(2-3p)/4}
\nu^{-p/2}$ ($t > t_{\rm m}$) for the high frequency light curve, and
$F_\nu\propto t \nu^2$ ($t < t_{\rm a}$), $F_\nu\propto t^{1/6}
\nu^{1/3}$ ($t_{\rm a} < t < t_{\rm 0}$), $F_\nu\propto t^{1/2}
\nu^{1/3}$ ($t_{\rm 0} < t < t_{\rm m}$), $F_\nu\propto t^{3(1-p)/4}
\nu^{-(p-1)/2}$ ($t_{\rm m} < t < t_{\rm c}$), and $F_\nu\propto 
t^{(2-3p)/4}\nu^{-p/2}$ ($t > t_{\rm c}$) for the low frequency 
light curve. The time and frequency dependences of the flux are 
sometimes parameterized 
using $F_\nu\propto t^\alpha \nu^\beta$. The relations between the
decay index $\alpha$ and spectral index $\beta$, $\alpha=3(1-p)/4
=3\beta/2$ ($t_{\rm m} < t < t_{\rm c}$ or $\nu_{\rm m} < \nu < \nu_{\rm 
c}$) and $\alpha=(2-3p)/4=(1+3\beta)/2$ ($t > t_{\rm c}$ or $\nu > 
\nu_{\rm c} > \nu_{\rm m}$) in the most relevant, slow cooling case,  
are often used to interpret optical and X-ray afterglow observations 
at relatively late times (of order one day or more). In the frequency 
regime $\nu < \nu_{\rm a,0}$, more relevant to radio afterglows, one 
has the ordering $t_0 < t_{\rm m} < t_{\rm a} < t_{\rm c}$, which 
yields $F_\nu\propto t \nu^2$ ($t < t_0$), $F_\nu\propto t^{1/2}
\nu^2$ ($t_0 < t < t_{\rm m}$), $F_\nu\propto t^{5/4}
\nu^{5/2}$ ($t_{\rm m} < t < t_{\rm a}$), $F_\nu\propto t^{3(1-p)/4}
\nu^{-(p-1)/2}$ ($t_{\rm a} < t < t_{\rm c}$), and $F_\nu\propto 
t^{(2-3p)/4}\nu^{-p/2}$ ($t > t_{\rm c}$). We reiterate that the 
scalings involving self-absorption in the fast cooling case with 
$t < t_0$ are derived assuming that the energy distribution of the 
cooling electrons is spatially homogeneous in
the emission region behind the shock front. The opposite situation 
where layers of cooling electrons at different distances from the 
shock front remain unmixed has been considered in detail by Granot 
et al. \cite{GPS00} 
 and Granot \& Sari \cite{GS01}. 

\subsection{Wind-Interaction Model}

The first modification of the standard model we consider is the density 
distribution
of the ambient medium. M\'esz\'aros, Rees, \& Wijers \cite{MRW98} studied 
the general case of a power-law ambient density distribution $n\propto
r^{-s}$, with an arbitrary power-index $s$.
Chevalier \& Li  \cite{CL99,CL00} 
examined the specific case of $s=2$, corresponding to a constant-mass 
loss rate, constant-velocity, circumstellar wind, possibly of a 
Wolf-Rayet origin. Some features of the wind-interaction model are
also described in Dai \& Lu \cite{DL98} and Panaitescu, M\'esz\'aros \& Rees \cite{PMR98}.
 Here, we follow the formalism of Chevalier \& Li \cite{CL00}. 

The wind-interaction model has the same instantaneous afterglow spectra
as in the standard ISM-interaction model, but different light curves. 
The differences in light curve come from the blast wave dynamics. 
For a relativistic blast wave propagating in an $s=2$ medium, Blandford 
\& McKee \cite{BM76}
 showed that its Lorentz factor $\gamma\propto  
R^{-1/2}$ in the adiabatic regime, and $\gamma\propto R^{-1}$ in the
fully radiative regime. Concentrating on the adiabatic evolution as
before and making the standard assumption about the 
fractions of the blast wave energy going into the electrons and magnetic 
fields, one finds that the characteristic frequencies scale 
with time as $\nu_{\rm a}\propto t^{-3/5}$, $\nu_{\rm m}\propto t^{-3/2}$, 
and $\nu_{\rm c}\propto t^{1/2}$, and the peak flux $F_{\nu,{\rm max}}
\propto t^{-1/2}$ in the slow cooling case. In the fast cooling case, the 
scalings are the same except for the self-absorption frequency, which 
now decreases with time more rapidly as $\nu_{\rm a}\propto t^{-8/5}$. 

The above scalings of the characteristic frequencies define four critical 
times, with four corresponding critical frequencies which divide the 
light curves into five distinct frequency regimes. For the typical 
parameters adopted 
by Chevalier \& Li \cite{CL00}, the light curve that is most relevant to the 
optical and X-ray afterglows has the ordering $t_{\rm a} < t_{\rm m} < 
t_0 < t_{\rm c}$, which yields $F_\nu\propto t^{7/4}\nu^{5/2}$ for $t 
< t_{\rm a}$, $F_\nu\propto t^{-1/4}\nu^{-1/2}$ for $t_{\rm a} < t < 
t_{\rm m}$, $F_\nu\propto t^{(2-3p)/4}\nu^{-p/2}$ for $t_{\rm m} < t < 
t_0$, $F_\nu\propto t^{(2-3p)/4}\nu^{-p/2}$ for $t_0 < t < t_{\rm c}$, 
and $F_\nu\propto t^{(1-3p)/4}\nu^{-(p-1)/2}$ for $t > t_{\rm c}$. Note
that as the cooling frequency $\nu_{\rm c}$ moves across the observing 
frequency $\nu$ {\it from below}, the light curve steepens from $\alpha
=(2-3p)/4$ to $(1-3p)/4$ by a modest amount $\Delta\alpha=1/4$. The 
same amount of steepening, from $\alpha=3(1-p)/4$ to $(2-3p)/4$, also
occurs in the standard ISM-interaction model, as $\nu_{\rm c}$ moves 
across $\nu$ {\it from above}. A signature of the wind-interaction model 
is the relatively fast decline of the light curve in the non-cooling 
spectral region $\nu_{\rm m} < \nu < \nu_{\rm c}$, where $\alpha=(1-3p)
/4$ compared with $3(1-p)/4$ for the ISM case. If the cooling frequency
falls between optical and X-ray wavelengths, the wind model would predict an X-ray 
light curve declining less steeply than an optical light curve, whereas 
the opposite would be true for the ISM model. 

At the much lower, radio frequency, the characteristic times typically 
have the following ordering: $t_{\rm c} < t_0 < t_{\rm a} < t_{\rm m}$,
which leads to a light curve with $F_\nu\propto t^{7/4}\nu^{5/2}$ for
$t < t_{\rm c}$, $F_\nu\propto t^2\nu^2$ for $t_{\rm c} < t < t_0$,
$F_\nu\propto t\nu^2$ for $t_0 < t < t_{\rm a}$, $F_\nu\propto t^0
\nu^{1/3}$ for $t_{\rm a} < t < t_{\rm m}$, and $F_\nu\propto t^{(1-3p)
/4}\nu^{-(p-1)/2}$ for $t > t_{\rm m}$. In the slow cooling case with 
$t > t_0$ most relevant to radio observations, the fairly steep 
rise of radio flux $F_\nu\propto t$ at the self-absorbed frequencies 
could in principle distinguish the wind model from the standard ISM 
model, where the rise is slower ($F_\nu\propto t^{1/2}$). In practice,
the difference is masked to a large extent by interstellar scintillation.

\subsection{Transition to Non-relativistic Evolution} 

After a GRB blast wave sweeps up a mass equivalent to the rest mass of 
the explosion, its evolution  becomes non-relativistic. In a medium 
of constant density $n$, this happens at a radius of order $R_{\rm 
nr}=1.2\times 10^{18} E_{52}^{1/3} n^{-1/3}$~cm where $E_{52}$ is the 
explosion energy in units of $10^{52}$~ergs and $n$ the ambient number 
density in units of cm$^{-3}$. The transition occurs on a time scale 
$t_{\rm nr}=R_{\rm nr}/c=1.2 (1+z) E_{52}^{1/3} n^{-1/3}$~yr (where
$z$ is the cosmological redshift), which is 
longer than the typical duration of GRB afterglow observations, unless 
the explosion energy is much lower than $10^{52}$~ergs \cite{FWK00}
 and/or the density much higher than 1~cm$^{-3}$  \cite{DL99}.
Provided that the minimum electron Lorentz factor 
$\gamma_{\rm m} > 1$, the temporal
and frequency dependences of the flux in the non-relativistic regime
can be obtained in a way similar to that in the relativistic regime 
\cite{WRM97}. Here, we follow the detailed 
treatment of Frail et al. \cite{FWK00}. The electrons are typically slow
cooling, with $\nu_{\rm m}\propto t^{-3}$, $\nu_{\rm c}\propto 
t^{-1/5}$, and the peak flux at $\nu_{\rm m}$ given by $F_{\nu,{\rm max}}
\propto t^{3/5}$. The self-absorption frequency could either be 
greater or smaller than $\nu_{\rm m}$. There are two relevant orderings
of characteristic frequencies: $\nu_{\rm a} < \nu_{\rm m} < \nu_{\rm c}$
and $\nu_{\rm m} < \nu_{\rm a} < \nu_{\rm c}$. In the former case, 
$\nu_{\rm a}\propto t^{6/5}$, one has  
$F_\nu\propto \nu^2 t^{-2/5}$ for $\nu < \nu_{\rm a}$, $F_\nu\propto 
\nu^{1/3} t^{8/5}$ for $\nu_{\rm a} < \nu < \nu_{\rm m}$, $F_\nu\propto
\nu^{-(p-1)/2} t^{3(7-5p)/10}$ for $\nu_{\rm m} < \nu < \nu_{\rm c}$,
and $F_\nu\propto \nu^{-p/2}t^{(4-3p)/2}$ for $\nu > \nu_{\rm c}$. 
In the latter case, $\nu_{\rm a}\propto t^{(2-3p)/(p+4)}$, and $F_\nu
\propto \nu^2 t^{13/5}$ for $\nu < \nu_{\rm m}$, $F_\nu\propto 
\nu^{5/2} t^{11/10}$ for $\nu_{\rm m} < \nu < \nu_{\rm a}$, $F_\nu\propto
\nu^{-(p-1)/2} t^{3(7-5p)/10}$ for $\nu_{\rm a} < \nu < \nu_{\rm c}$,
and $F_\nu\propto \nu^{-p/2}t^{(4-3p)/2}$ for $\nu > \nu_{\rm c}$. Note
that the light curve decreases with time faster in the
non-relativistic regime (where $\alpha=3[7-5p]/10$ and $[4-3p]/2$ before 
and after the cooling break $t_{\rm c}$, or $\alpha=-1.65$ and $-1.75$ 
for $p=2.5$) than in the relativistic regime (where the corresponding 
$\alpha=3[1-p]/4$ and $[2-3p]/4$ or $\alpha=-1.125$ and $-1.375$). 

In the wind interaction model with an ambient density $\rho=A r^{-2}$, 
the transition to non-relativistic evolution 
occurs at a distance $R_{\rm nr}=1.8\times 10^{18} E_{52}/A_*$~cm,
where $A_*$ is the coefficient $A$ in units of $5\times 10^{11}$~g~cm$^{-1}$
\cite{CL00}, corresponding to a time $t_{\rm nr}=1.9(1+z)E_{52}
/A_*$~yr, again longer than most afterglow observations for typical 
parameters. In the non-relativistic regime, one finds $\nu_{\rm m}\propto 
t^{-7/3}$, $\nu_{\rm c}\propto t$, and the peak flux $F_{\nu,{\rm 
max}}\propto t^{-1/3}$ at $\nu_{\rm m}$. The relevant orderings are 
$\nu_{\rm a} < \nu_{\rm m} 
< \nu_{\rm c}$ and $\nu_{\rm m} < \nu_{\rm a} < \nu_{\rm c}$. In the former 
case, $\nu_{\rm a}\propto t^{13/15}$, and   
$F_\nu\propto \nu^2 t^{-1}$ for $\nu < \nu_{\rm a}$, $F_\nu\propto 
\nu^{1/3} t^{4/9}$ for $\nu_{\rm a} < \nu < \nu_{\rm m}$, $F_\nu\propto
\nu^{-(p-1)/2} t^{(5-7p)/6}$ for $\nu_{\rm m} < \nu < \nu_{\rm c}$,
and $F_\nu\propto \nu^{-p/2}t^{(8-7p)/6}$ for $\nu > \nu_{\rm c}$. 
In the latter case, $\nu_{\rm a}\propto t^{(4-7p)/[3(p+4)]}$, and $F_\nu
\propto \nu^2 t^{-1}$ for $\nu < \nu_{\rm m}$, $F_\nu\propto 
\nu^{5/2} t^{1/6}$ for $\nu_{\rm m} < \nu < \nu_{\rm a}$, $F_\nu\propto
\nu^{-(p-1)/2} t^{(5-7p)/6}$ for $\nu_{\rm a} < \nu < \nu_{\rm c}$,
and $F_\nu\propto \nu^{-p/2}t^{(8-7p)/6}$ for $\nu > \nu_{\rm c}$. As
in ISM interaction, the light curve decreases with time faster in the
non-relativistic regime (where $\alpha=[8-7p]/6$ and $[5-7p]/6$ before 
and after the cooling break $t_{\rm c}$, or $\alpha=-1.58$ and $-2.08$ 
for $p=2.5$) than in the relativistic regime (where the corresponding 
$\alpha=[2-3p]/4$ and $[1-3p]/4$ or $\alpha=-1.375$ and $-1.625$). 

\subsection{Jet Model}

Starting with Rhoads \cite{R97}, jet models of afterglows have been widely 
discussed in connection with the steepening of light curves. We will 
first outline the asymptotic analysis of Sari, Piran \& Halpern \cite{SPH99} 
and then comment on possible complications. Recent reviews of the
subject include \cite{R01} and \cite{PG01}.

Let $\theta_0$ be the initial angular width of the jet. When the Lorentz 
factor $\gamma$ drops below $\theta_0^{-1}$, the jet starts to spread 
sideways, changing its dynamics. The spreading occurs around a time
$t_{\rm jet}=3[(1+z)/2](E_{52}/n)^{1/3}(\theta_0/0.2)^{8/3}$~days in
a constant-density medium and $t_{\rm jet}=2[(1+z)/2](E_{52}/A_*)
(\theta_0/0.2)^4$~days in a wind \cite{CL00}. Before $t_{\rm 
jet}$, the standard spherical results apply. After $t_{\rm jet}$, the
jet Lorentz factor decreases with distance exponentially \cite{R97},
 yielding $\gamma\propto t^{-1/2}$, $\nu_{\rm 
m} \propto t^{-2}$ and $\nu_{\rm c}\propto t^0$, and the 
peak flux $F_{\nu,{\rm max}}\propto t^{-1}$ at $\nu_{\rm m}$. In the
case $\nu_{\rm a} < \nu_{\rm m} < \nu_{\rm c}$, one finds that 
$\nu_{\rm a}\propto t^{-1/5}$, and $F_\nu\propto \nu^2 t^0$ for 
$\nu < \nu_{\rm a}$, $F_\nu\propto 
\nu^{1/3} t^{-1/3}$ for $\nu_{\rm a} < \nu < \nu_{\rm m}$, 
$F_\nu\propto \nu^{-(p-1)/2} t^{-p}$ for $\nu_{\rm m} < \nu < 
\nu_{\rm c}$, and $F_\nu\propto \nu^{-p/2}t^{-p}$ for $\nu > 
\nu_{\rm c}$. In the case $\nu_{\rm m} < \nu_{\rm a} < \nu_{\rm c}$, 
we have instead $\nu_{\rm a}\propto t^{-2(p+1)/(p+4)}$, and $F_\nu
\propto \nu^2 t^0$ for $\nu < \nu_{\rm m}$, $F_\nu\propto 
\nu^{5/2} t$ for $\nu_{\rm m} < \nu < \nu_{\rm a}$, $F_\nu\propto
\nu^{-(p-1)/2} t^{-p}$ for $\nu_{\rm a} < \nu < \nu_{\rm c}$, and 
$F_\nu\propto \nu^{-p/2}t^{-p}$ for $\nu > \nu_{\rm c}$. These
scalings apply to both ISM and wind interaction models. 

The steepening of light curves to an asymptotic scaling of $F_\nu
\propto t^{-p}$ is a signature of the jet models. How sharply the 
change from one temporal slope to another actually happens remains 
controversial. M\'esz\'aros \& Rees \cite{MR99} pointed out that seeing
the edge of a jet when its Lorentz factor drops below $\theta_0
^{-1}$ would also steepen the light curve, by a factor of 
$\Delta\alpha=3/4$ in a constant-density medium and $\Delta\alpha
=1/2$ in a wind. Semi-analytic calculations 
taking into account of both the sideways spreading and edge effect 
find that the transition from one slope to another tends to be
continuous, spanning one decade or more in the observer's time 
\cite{MSB00,WL00,KP00a},
especially in a wind-type ambient medium. These calculations
adopted a simplified set of 1D equations for the jet dynamics, which 
need to be checked against numerical simulations. Preliminary 2D 
calculations of Granot et al. \cite{Ge01} show that the blast wave has
an egg-like shape, differing considerably from that predicted from 
the analytic or semi-analytic models. A steepening of light curves 
does occur around the time $\gamma\propto \theta_0^{-1}$, as expected 
from simple arguments. Whether the sharpness and amount of steepening
match the predictions of the semi-analytic models remain to be seen. 

\subsection{Other Modifications} 

Many additional effects have been considered in the afterglow literature
on top of the basic interaction models outlined above. We will limit 
ourselves to three of the more commonly discussed ones: inverse Compton 
scattering, pair production, and a non-standard electron energy 
distribution. 

Inverse Comptonization of synchrotron photons by relativistic electrons 
could not only produce a high energy component of afterglow emission that 
is potentially observable but also dominate the cooling of electrons 
and affect the blast wave dynamics 
\cite{PM98,WL98,CD99,PK00}. Sari \& Esin \cite{SE01}  
 examined these effects in detail, and concluded that
the inverse Compton spectra broadly resemble the primary synchrotron
spectra in shape, although significant differences do exist, especially
at the high frequency end where a broken power-law description is 
no longer adequate. They showed that as long as the fraction of the 
blast wave energy 
in electrons $\epsilon_{\rm e}$ exceeds that in magnetic fields $\epsilon
_{\rm B}$, the inverse Compton emission dominates the synchrotron 
emission in cooling the electrons in the fast cooling regime. Depending 
on the ratio $\epsilon_{\rm e}/\epsilon_{\rm B}$, the domination can 
extend well into the slow cooling regime, changing the value of the 
cooling frequency $\nu_{\rm c}$ and thus the light curves of afterglow 
emission. 

Thompson \& Madau \cite{TM00} considered pair formation as the gamma-ray 
photons from the GRB proper propagate ahead of the external shock 
front and interact with the seed photons back-scattered by the ambient 
medium. A simplified discussion of the process is presented in 
M\'esz\'aros, Ramirez-Ruiz \& Rees \cite{MRR01}. Beloborodov \cite{B01} went 
one step further and solved for the dynamics of the pair-loaded medium. 
These studies concluded that pair production can dramatically increase 
the radiative efficiency of the blast wave and potentially broaden the 
original pulses of gamma-rays if the ambient density is high enough. 
The effects on afterglows are expected to be large at early times, 
especially for the wind-interaction model, although details are yet to 
be worked out. 

Most afterglow models assume that the radiating electrons have a power-law 
energy distribution at injection above some cutoff. The power-law indexes 
inferred from afterglow observations under this assumption span a wide 
range, from $p \sim 2$ (or below) to $\sim 3$ (e.g., \cite{CL00,PK01a,PK01b}).
 The lack of a universal value for $p$ calls
into question the assumption that the shock front accelerates electrons
to a power-law that is constant with energy and with time. Indeed, in the 
Crab Nebula, arguably the best studied astrophysical synchrotron source, 
a break in the injection spectrum is required \cite{Ae00}. 
This motivated Li \& Chevalier \cite{LC01} to consider a broken power-law 
distribution of electron energy that steepens at high energies in the 
context of the spherical 
wind-interaction model. Similar non-standard energy distributions have 
been adopted in  other models \cite{DC01,P01,WL01}. 
A major attraction of the invoked steepening
of electron energy distribution is that it can lead to a steepening of 
afterglow light curves. The light curve steepening is chromatic, and 
should be distinguishable from the achromatic steepening due to jet 
effects if wide wavelength coverage is available. 

\section{Application to Individual Sources}
\label{application}

We now apply the interaction models to the dozen or so GRBs whose afterglows
are relatively well observed. It turns out that for most of the sources 
the models are not unique. We shall start the discussion with the cases 
that in our view are less controversial. 

\subsection{Probable Collimated ISM Interactors: GRBs 990123 and 990510}

The first two GRBs that show breaks in their optical afterglows are GRB 
990123 \cite{Ke99} and GRB 990510 \cite{He99}. 
The breaks in both sources are achromatic and are interpreted as due 
to jets. In the case of GRB 990123, the break occurs at a time $t_{\rm 
b}=1.68\pm 0.19$~days, 
depending somewhat on the fitting function adopted  
 \cite{Ho01}. Before the break, the temporal decay index at optical wavelengths 
is $\alpha=-1.12\pm 0.08$. The optical spectral index is found to be 
$\beta=-0.750\pm 0.068$. These two indexes can be fitted by an 
ISM-interaction model with $p=2.5$ in the adiabatic regime (where
$\alpha=-3[p-1]/4$ and $\beta=-[p-1]/2$ for $\nu_{\rm m} < \nu < 
\nu_{\rm c}$). The decay index $\alpha=-1.44\pm 0.07$ at X-ray wavelengths 
is also consistent with the model, provided that the X-rays are in 
the cooling regime (where $\alpha=[2-3p]/4$ for $\nu > \nu_{\rm c}$).
The observed steeper decline in X-rays than in optical is expected 
for ISM interaction but not for wind interaction \cite{CL99}.
After the break, the optical light curves steepen quickly to 
$\alpha=-1.69\pm 0.06$ while the spectral index $\beta$ remains 
approximately the same  \cite{Ho01}. The observed amount of 
steepening, $\Delta\alpha=0.57\pm 0.10$, agrees marginally with that 
expected of a jet with a fixed opening angle (where $\Delta\alpha=3/4$ 
for ISM interaction). It is less than half of the asymptotic value 
$\Delta\alpha=(p+3)/4=1.38$ expected from the sideways expansion of a 
jet. The relatively small amount of steepening was taken as evidence 
for seeing the edge of a non-spreading jet  \cite{MR99,Ke99,Ho01}, 
although it does not rule 
out jet-spreading as the cause of steepening, since it takes time for 
the light curves to reach the asymptotic slope. Indeed, Panaitescu \& 
Kumar  \cite{PK01a} obtained a
reasonable fit to the multi-frequency data of GRB 990123 using a 
semi-analytic jet model taking into account of lateral expansion. 
%% further support from reverse shock?

The breaks in the optical light curves of GRB 990510 appear to be
smoother than those in GRB 990123. Fitting a continuous function,
Harrison et al.  \cite{He99} obtained a break time of $t_{\rm b}=1.20\pm
0.08$~days. Before the break, the decay index $\alpha=-0.82\pm 0.02$, 
which implies $p=2.1$ in the adiabatic regime for ISM-interaction. 
The expected spectral index $\beta=-0.55$ is consistent with $\beta=
-0.61\pm 0.12$ determined by Stanek et al.  \cite{Se99} or $\beta=-0.531
\pm 0.019$ by Holland et al.  \cite{Ho01}. The latter authors find some 
evidence that $\beta$ decreases with time which, if real, is not 
explained. The relatively flat decay rate of $\alpha > -1$ is 
difficult to accommodate in the wind-interaction model; it requires 
a very flat electron energy distribution with $p < 2$ \cite{CL00}. 
Even then, the predicted spectral
index would not match that observed  \cite{PK01a}. After 
the break, the measured $\alpha=-2.18\pm 0.05$, which is consistent
with the expected asymptotic value $\alpha=-p$  \cite{SPH99}. 
The ISM-interacting jet model is further supported by radio data,
which are consistent with the expected $F_\nu\propto t^{-1/3}$
evolution  \cite{He99}. X-ray data are available between
0.3 and 2 days, bracketing the break in optical  
 \cite{Ku00}. The lightcurve can be fitted either by single power-law with 
$\alpha=-1.42\pm 0.07$ or a broken-power law, and is well modeled 
using an ISM-interacting jet model with the cooling frequency 
$\nu_{\rm c}$ between the optical and X-rays  
 \cite{PK01a,Pie01}. 
The detection of polarization from this event
 \cite{Ce99,We99} is also consistent with 
the jet interpretation  \cite{GL99,S99}, although
the observed polarization is small and does not require a jet. 

%% Alternative to jet model for GRB 990123 or GRB 990510?

\subsection{Probable Spherical Wind Interactors: GRBs 980425 and 970508}

The probable association of GRB 980425 with the radio supernova SN 1998bw 
at $z=0.0085$ is discussed in the Chapters by Galama and Iwamoto \& Nomoto. 
Here, we are concerned with the radio emission from this source, which is 
extraordinary \cite{Ke99}. Kulkarni et al. \cite{Ke99} argued that the shock 
responsible for the radio emission is relativistic, based on the high 
brightness temperature, on a synchrotron self-absorption interpretation 
of the early evolution, and on scintillation results. Using the standard 
synchrotron theory, Li \& Chevalier \cite{LC99} were able to deduce from the 
radio data that the shock is expanding into a medium with an approximately 
$r^{-2}$ density profile at a speed comparable to the speed of light. 
A more detailed model, taking into account of the shock dynamics and 
relativistic effects and assuming the standard power-law electron
distribution above some cutoff, reproduces the radio lightcurves after 
about day 10 reasonably well, provided that the shock energy is increased 
by a factor of $\sim 2.5$ at $\sim 10^2$ days in the rest frame of the 
explosion to explain the rise in the radio fluxes observed between days 
20--40. Based on the radio spectra at day 12 and day 15 and adopting
a mono-energetic distribution for the synchrotron-emitting electrons, 
Waxman \& Loeb \cite{WL99} inferred a shock speed of $\sim 0.3$~c, which is a 
factor of $\sim 2$ smaller than that inferred by Li \& Chevalier \cite{LC99} 
around the same time. The longer term evolution of the radio source
implies a power law particle energy distribution, as generally
observed in GRB afterglows.  Weiler, Panagia \& Montes \cite{WPM01} were able to 
fit a  parameterized model to the radio data, and 
came to a similar conclusion that the shock is mildly relativistic and 
is interacting with a clumpy progenitor wind.  

Chevalier \& Li \cite{CL00} applied the spherical wind-interaction model to
GRB 970508 which, unlike GRB 980425, is at a cosmological distance 
of $z=0.835$  \cite{Me97}. Extensive optical data up to 
hundreds of days are available for this source (see Chapter by 
Pian \& Fruchter). They follow a power-law after about day 2 with 
$\alpha=-1.141\pm 0.014$ and $\beta=-1.11\pm 0.06$  \cite{Ge98a}. 
The X-rays have a similar decay index of $\alpha=-1.1\pm 0.1$
 \cite{Pe98} and the optical/X-ray spectral index is consistent
with $\beta=-1.1$  \cite{Ge98a}. Based on the radio to X-ray 
spectrum on day 12.1, Galama et al.  \cite{Ge98b} deduced that the cooling 
frequency is just below optical frequencies. At the optical and X-ray frequencies
above the cooling frequency $\nu > \nu_{\rm c}$, the observed value
of $\alpha$ implies $p=2.2$, which yields a spectral index $\beta$
consistent with that observed for both the ISM and wind interaction 
models. Radio observations are crucial for distinguishing between 
these two models. In Fig.~1(a), we show the predictions of a wind 
interaction model at three wavelengths. They fit the radio data
reasonably well, especially after about day 100 when the interstellar 
scintillation dies down  \cite{FWK00}. In Fig.~1(b), we present 
an analytic fit to the R-band data, which is used in conjunction 
with the radio fit to obtain the best model parameters: $\epsilon_{\rm 
e}=0.2$, $\epsilon_{\rm B}=0.1$, $E_{52}=0.3$, and $A_*=0.3$. The 
inferred mass loss rate of $3\times 10^{-6}$~M$_\odot$~yr$^{-1}$ (for 
a wind speed of $10^3$~km~s$^{-1}$) is in the expected range of a 
Wolf-Rayet star. The optical afterglow brightens unexpectedly around 
day 2. Its temporal and spectral behaviors before day 2 are not explained 
by the wind model. 

\begin{figure}%[b]
\begin{center}
\includegraphics[width=.6\textwidth]{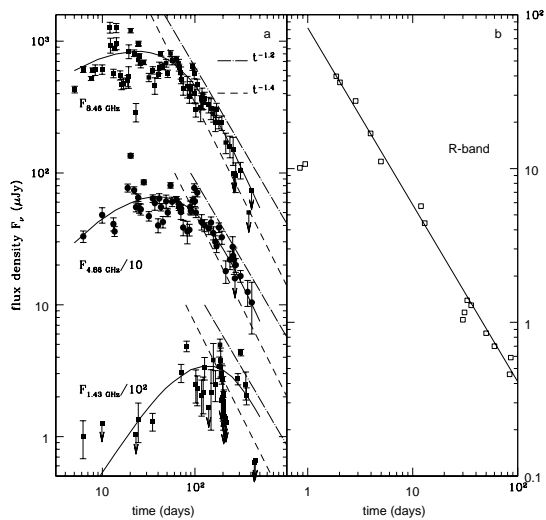}
\end{center}
\caption[]{Wind interaction model for the afterglow of GRB 970508 (adopted
from Chevalier \& Li \cite{CL00}). Radio data are taken from Frail et al. 
\cite{Fe97,FWK00}
 and R-band data from Sokolov et al. \cite{Se98}.}  
\label{fig1}
\end{figure}

The peak flux $F_{\nu_{\rm m}}$ and absorption frequency $\nu_{\rm a}$ 
of GRB 970508 are inferred to decrease with time, which is consistent 
with the wind model but not with the standard spherical ISM model in 
the relativistic regime  \cite{Ge98a}. Frail et al. 
\cite{FWK00} discussed in depth these problems and suggested jet effects and 
a transition to non-relativistic expansion as possible solutions. They
proposed three phases of evolution in a constant-density medium: a 
relativistic jet phase followed by a phase of jet spreading, which is
in turn followed by a phase of non-relativistic, spherical expansion. 
The jet spreading reduces the peak flux $F_{\nu_{\rm m}}$, and the 
emission expected from the non-relativistic phase is shown to be 
consistent with the radio data, although the complete model has yet 
to be calculated. 
The relatively slow decline of the optical afterglow for over 100 days \cite{Fre00}
was unusual compared to other well-studied afterglows.

\subsection{Jets, Winds, or Non-Relativistic Evolution?}

Besides the four sources discussed above, there are half-dozen or so GRBs
with reasonable multi-frequency coverage of afterglows, particularly
at radio frequencies, which allows for detailed modeling. These include 
GRBs 980519, 991208, 991216, 000301C, 000418, 000926, and 010222 as of 
August 2001. The majority of these sources show clear steepening in the
optical light curves (GRBs 991216, 000301C, 000926, and 010222), while 
in others the evidence for light curve steepening is weaker (GRBs 980519 and 
991208) or absent (GRB 000418). A variety of models has been proposed
for these sources. One common conclusion is that the simplest, standard 
model involving a relativistic spherical blast wave expanding into a 
constant-density medium is inadequate, and further refinements are  
necessary. GRB 000418 was modeled by Berger et al.  \cite{Be01b}, who 
found that 
both a spherical wind model and an ISM-interacting jet model fit the
available radio-to-optical data reasonably well. For the remaining 
six sources, Panaitescu \& Kumar  \cite{PK01a,PK01b,PK01c} constructed 
semi-analytic 
jet models for both constant-density and wind-type ambient media. They 
found that ISM-interacting jet models can plausibly explain the 
broadband emission of all sources, and that wind-interacting 
jet models are compatible with the emission of GRBs 991208 and 991216 
and can marginally accommodate the afterglows of GRBs 000301C and 010222 
but cannot explain the observations of GRBs 980519 and 000926. Other 
authors have modeled these sources individually and, in some cases, come 
to different conclusions. We discuss these sources in turn.

GRB 980519 is one of three sources that show unusually steep light curve 
decay (with $\alpha < -2$) but weak or no evidence for breaks (the other 
two being GRBs 980326 and 991208). Chevalier \& Li  \cite{CL99} proposed that
the steep decay can be explained in a spherical wind model, provided
that the electron energy index $p$ is close to 3, a value that is higher
than normally found in GRB afterglows but is within the range found in
radio supernovae \cite{C98}. Radio data provide some support to
this interpretation  \cite{Fe00a} but do not rule out the jet model, 
first advocated for this source by Sari et al. \cite{SPH99}.  Indeed, Jaunsen 
et al. \cite{Je01} have recently found some evidence for a sharp break in
the R-band light curve, which they interpreted as due to a jet expanding 
into a wind-type medium. The relative sparseness of the data available 
around and after the break makes its identification less secure. If true, 
the sharpness of the break would be difficult to understand in a jet 
model, especially if the ambient medium is wind-like  
 \cite{KP00a}. The spherical wind model of Chevalier \& Li \cite{CL99} 
did not take 
into account of inverse Compton scattering, which is important for 
the parameters they adopted (e.g., \cite{SE01}). Inverse Compton 
scattering lowers the cooling frequency, and poses a problem for the 
wind model in fitting the X-ray afterglow. 

GRB 991208 is unique in that the radio data are sampled well enough to
allow for a determination of the evolution of the characteristic
frequencies $\nu_{\rm a}$ and $\nu_{\rm m}$ and the peak flux $F_{\nu,
{\rm max}}$ at $\nu_{\rm m}$. The inferred scalings $\nu_{\rm a}
\propto t^{-0.15\pm 0.23}$, $\nu_{\rm m}\propto t^{-1.7\pm 0.7}$
and $F_{\nu,{\rm max}}\propto t^{-0.47\pm 0.20}$   \cite{Ge00b}
are compatible with the predictions of either the ISM-interacting jet 
model ($\nu_{\rm a}\propto t^{-1/5}$, $\nu_{\rm m}\propto t^{-2}$,
and $F_{\nu,{\rm max}}\propto t^{-1}$) or the spherical wind model
($\nu_{\rm a}\propto t^{-3/5}$, $\nu_{\rm m}\propto t^{-3/2}$,
and $F_{\nu,{\rm max}}\propto t^{-1/2}$) but not with those of the 
spherical ISM model ($\nu_{\rm a}\propto t^0$, $\nu_{\rm m}\propto 
t^{-3/2}$, and $F_{\nu,{\rm max}}\propto t^0$). Galama et al.  \cite{Ge00b}
pointed out a problem with the wind model: the rate of optical 
light curve decline implies $p\ge 3.3$ and such a large value of $p$ 
would be inconsistent with the radio-to-optical spectrum at day 7.3.
The inconsistency motivated Li \& Chevalier \cite{LC01} to seek a 
model with a
non-standard energy distribution of electrons. They found that a
spherical wind model with a broken power-law electron energy 
distribution can reproduce all data as well as, if not better than, 
the jet model. Interestingly, in the jet model of Panaitescu \&
Kumar  \cite{PK01b} for this source, where the jet dynamics is followed
semi-analytically, the steep decay in optical is mostly attributed   
to a steepening in the electron energy distribution, as in the wind
model for this source, rather than jet effects. 

GRB 000301C was modeled by Li \& Chevalier \cite{LC01} using the same 
spherical wind model with an electron energy distribution that 
steepens at a certain (high) energy as the one applied to GRB 991208. 
The quality of the overall fit to the broad band radio-to-optical 
data is again comparable to that of an ISM-interacting jet model 
 \cite{Be00}. The radio data at 8.48 GHz are particularly 
well sampled for this source. The fact that its temporal decay 
(with $\alpha\approx -1.4$ \cite{Be00}) is much shallower 
than that observed in R-band (with an asymptotic value of $\alpha
\approx -2.7$) provides some evidence for a steepening of the 
electron energy distribution at high energies (see also \cite{PK01b}). 
The steepening in
energy distribution produces a steepening in the optical light curves, 
which could mimic the jet effects. A difference is that the optical 
spectrum should steepen with time in the former case but not in the 
latter. Rhoads \& Fruchter  \cite{RF01} presented IR-optical-UV data at
several epochs, showing some evidence for spectral steepening 
with time. These data are fitted reasonably well by the wind model. 
The spectral steepening would be more pronounced in the optical to 
X-ray regime. Unfortunately, X-ray observations are not available
for this source. Kumar \& Panaitescu  \cite{KP00b} attributed the steepening 
of the R-band light curve to a sudden large drop in the density of 
the ambient medium and Dai \& Lu  \cite{DL01} to transition to 
non-relativistic 
evolution. The predictions of these models on radio emission remain 
to be worked out and compared with observations. 

GRB 991216 is a source well observed in radio, optical and X-rays. The 
decay indexes at these wavelengths ($\alpha=-0.82\pm 0.02$ at 8.46 GHz, 
$-1.33\pm 0.01$ at R-band and $-1.61\pm 0.06$ at 2--10 kev) are all 
different  \cite{Fe00b}. There is some evidence for light curve
steepening at R-band, although the data at late times are too sparse
to tightly constrain the time of transition from one power-law decay to 
another 
and the decay index after the transition  \cite{He00}. 
The optical to X-ray data are consistent with an ISM-interacting jet
model \cite{He00}, but the radio data are not 
 \cite{Fe00b}. Frail et al. \cite{Fe00b} attributed the bulk of the 
radio emission to either 
the reverse shock as in GRB 991023  \cite{SP99} or a second 
forward shock. Panaitescu \& Kumar  \cite{PK01a} examined the multi-frequency 
data, and concluded that all of the data can be fitted reasonably well
using an ISM-interacting jet model, provided that a large curvature 
exists in the electron energy distribution, with $p$ changing from 
$\sim 1.2$ to $\sim 2.1$. GRB 991216 is therefore the third source 
(after GRBs 991208 and 000301C discussed above; see also  
 \cite{PK01b}) for which a steepening in the electron energy distribution 
is proposed, although in all three cases the interpretation is model 
dependent. 

GRB 000926 stands apart from other sources in that its X-ray afterglow 
is observed nearly two weeks after the explosion  
\cite{Pire01}.
There is clear evidence for steepening in the optical light curves,
with the decay index changing from $\alpha=-1.46\pm 0.11$ to $-2.38
\pm 0.07$ around $1.8\pm 0.1$ days  \cite{He01,Pre01,Fe01,Se01a}, 
pointing to a jet 
explanation. However, the decay after the break is significantly 
steeper than that in X-rays, estimated to be $\alpha=-1.89^{+0.19}
_{-0.16}$  \cite{Pire01}, which is difficult to explain in the
standard jet model. Piro et al.  \cite{Pire01} proposed an alternative 
model involving a mildly collimated blast wave expanding into a dense 
uniform medium ($n\sim 3\times 10^4$~cm$^{-3}$), with the transition 
to non-relativistic evolution 
occurring $\sim 5$ days after the explosion. Harrison et al.  \cite{He01}
showed that such a high density appears to have difficulties 
reproducing the radio data. They suggested that the standard jet
model, either in a constant-density or wind-like medium, can fit 
the broad band data reasonably well, provided that the inverse
Compton emission contributes significantly to the X-ray emission (see 
also  \cite{Ye01} and \cite{PK01c}). 

GRB 010222 is well observed in X-rays, with a decay index $\alpha=-
1.33\pm 0.04$  \cite{ie01} that is nearly identical to 
that in the optical after a break around $\sim 0.5$~days (e.g., 
 \cite{Me01}). Before the break, the optical light curve 
is significantly flatter, with $\alpha\sim -0.6$ to $-0.8$. The 
steepening
of optical light curves can be interpreted as due to jet effects 
\cite{Se01,Se01b,Ce01}, 
provided that the electron energy distribution is very flat (with 
$p\sim 1.5$) as required by the asymptotic relation $p=-\alpha$
after the break  \cite{SPH99}. This relation, originally 
derived for $p > 2$, may not be applicable to the case of $p < 2$ 
\cite{B01,DC01}. 
Masetti et al.  \cite{Me01} and
in 't Zand et al.  \cite{ie01} favored an alternative model, in which 
the transition to non-relativistic evolution occurs rapidly in
a very dense medium ($n\sim 10^6$~cm$^{-3}$). The inferred 
$p\approx 2.2$ is more in line with those inferred for other GRBs 
(see however  \cite{PK01b}). It is not clear, however, 
whether this high density model can reproduce the early radio 
detection of the source \cite{Be01a}.

\section{Discussion and Conclusions}
\label{discussion}

No single model explains all of the dozen or so relatively well 
observed GRB afterglows. The model that comes closest appears 
to be the ISM-interacting jet model at the present time. It fits 
particularly well the 
multi-frequency data of GRBs 990123 and 990510. To reproduce the 
afterglow observations of GRBs 991208, 991216 and 000301C, a
relatively flat electron energy distribution with $p < 2$ is 
required; the flat distribution must steepen at 
high energies to $p > 2$ for the total electron energy to remain 
bound. The model also appears  
capable of describing the data of GRBs 000418  \cite{Be01b}
and 000926, provided that inverse Compton scattering 
contributes significantly to the X-ray emission of the latter 
 \cite{He01}. In addition, the 
steep decay of the optical light curves of GRBs 980326 and 
980519 has been attributed to jet effects  \cite{SPH99},
and GRB 970508 has been modeled as a mildly collimated jet 
making a transition to a spherical non-relativistic evolution
 \cite{FWK00}. 

However, a number of the above sources 
can be modeled equally well or perhaps better by the wind-interacting 
spherical 
model. This is particularly true for GRB 970508. In the case of GRBs 
991208 and 000301C, a non-standard electron energy distribution
is required \cite{LC01}, as is the case for GRB 
991216 (Li \& Chevalier, in preparation). In addition, the 
wind model can fit the extensive radio data of GRB 980425/SN 1998bw, 
provided that a late energy injection occurs. Furthermore, the optical 
and X-ray afterglow data of GRB 970228 are compatible with the wind 
model, after the subtraction of plausible supernova emission
from the optical light curve \cite{Rei99,Ge00a}. 
Frail et al.  \cite{Fe98} searched for this source in radio for the first
year, and did not detect any afterglow. The upper limits are of 
order 100 $\mu$Jy or less at 8.46 GHz where the monitoring is most 
frequent. This absence of radio afterglow is not easy to explain 
in the standard, ISM-interacting spherical model, but is 
compatible with either a spherical wind model or ISM-interacting 
jet model \cite{CL00}, since in both models the 
peak flux decreases with frequency. The jet model may have difficulty 
with the relatively flat decay of the optical light curve ($\alpha\sim 
-1.6$) unless the electron energy distribution is very flat and/or
the asymptotic slope after the jet break has yet to be reached. 
A relatively complete optical/IR data set also exists for GRB 980703 
 \cite{Be98,CTe99,Ho01}. The inferred 
large extinction intrinsic to its bright host galaxy, coupled
with the sparseness of the published X-ray \cite{Ve99}
and radio data \cite{Be98}, makes it impossible to constrain 
models firmly. 

The presence or absence of supernova emission in the optical afterglow 
may provide a powerful means for corroborating the ISM or wind model. 
The ISM-interacting sources are not expected to be accompanied by 
supernovae, and one indeed finds no evidence for supernova in the most 
probable ISM interactors: GRBs 990123 and 990510. The wind-interacting 
sources, on the other hand, are expected to be associated with supernovae,
and the best wind interactors GRBs 980425 and 970508 are probably 
associated with supernovae. The evidence of supernova association 
is strong for GRB 980425, as reviewed by Galama and Iwamoto \& Nomoto
in this book. The evidence is weaker for GRB 970508, based mainly on
a ``shoulder'' in the late time light curve at the $I_c$-band 
 \cite{S01}; supernova emission is not clearly seen in the R-band 
\cite{Fre00}.
% implying that a supernova in GRB 970508 would have 
%to be about 1 mag fainter than SN 1998bw. 
In the case of another possible 
wind interactor, GRB 970228,
 evidence for supernova has been marshaled by Reichart  \cite{Rei99}
and Galama et al.  \cite{Ge00a}; the alternative explanation involving dust 
echoes  \cite{EB00}
 appears less likely  \cite{Rei01}, 
especially if the GRB progenitors are Wolf-Rayet stars instead of 
red supergiants  \cite{C01}.
 Association with a supernova has also 
been proposed for other sources, including GRBs 980326 
 \cite{Be99}, 991208  \cite{CTe01},
 and possibly 000418  \cite{Ke00,DD00}. Their afterglow observations 
appear to be compatible with both the jet and wind models. 
In no source can the presence of a supernova be 
excluded definitively 
\cite{Fe01,DDR01}. 

The expansion into a wind-type medium and the association with
supernova would point to a massive star origin for GRBs. The
massive star connection is indirectly supported by the analysis
of optically dark bursts by Reichart  \cite{Rei01b}, which indicates 
that the majority of GRBs may be tied to giant molecular clouds, 
the sites of massive star formation. The connection may be strengthened
by the presence of possible Fe line emission in the X-ray afterglows 
of GRBs 970508 \cite{Pe99},
 970828 \cite{Ye99} 
%see however Djorgovski et al. 2001), 
991216  \cite{Pire00}, 
and 000214 \cite{Ane00}, which implies a substantial 
amount of circumburst material. We have discussed GRB 970508 
as a wind interactor. %% The afterglow of GRB 991216 can 
%be fitted by either a jet  \cite{PK01a} or a wind 
%model (Li \& Chevalier, in preparation). In both models, a 
%non-standard electron energy distribution must be invoked, however. 
%The remaining sources, 
GRBs 970828 and 000214, are optically dark,
which is compatible with the idea that they are tied to giant 
molecular clouds and thus (indirectly) to massive stars. Their 
afterglow data are insufficient to constrain ambient interaction 
models. 
However, some models for the Fe lines require a large Fe mass and
densities $\gsim 10^8$ cm$^{-3}$ at $r\approx 10^{16}$ cm
\cite{We00,Ve01}, much larger than the densities inferred in
afterglow models and than expected in the wind from a massive
progenitor star.
The distribution of the dense gas is asymmetric and it is not
present along the line of sight to the GRB, which is the region
probed by the afterglow emission.
The strong angular variation of the density is surprising;
in the supranova model \cite{Ve01}, the supernova would be 
expected to affect the density along the line of sight. The 
required Fe mass would be greatly reduced if the lines come 
from a much denser region closer to the explosion center,
as in the collapsar-bubble model of M\'esz\'aros \& Rees 
\cite{MR01}.

%The possible presence of supernova and Fe line emission in the
%optical and X-ray afterglows of GRB 970508 strengthens the case
%that it is a wind-interactor, which implies a massive star 
%progenitor. 
The absence of supernova emission in the afterglows of GRBs 
990123 and 990510 is consistent with their being ISM-interactors
with compact star merger progenitors. It therefore appears that 
there are two types of burst progenitors, with GRBs 970508 and 
980425 representing the best examples of one type, and GRBs 990123 
and 990510 the other
(\cite{CL00}, see also \cite{LW00}). The possible presence of supernova 
emission 
in the optical afterglows of GRBs 970228, 980326, 991208, and
000418,  % and the detection of Fe line emission in GRB 991216, 
coupled 
with the fact that their afterglows can be fitted by the wind model,
plausibly put them in the massive star category. This assignment 
is weakened, however, by the fact that the afterglows of these sources 
can be fitted by the ISM-interacting jet model as well. For the other 
GRBs with relatively well observed afterglows, the progenitor 
types are even less certain. 

The interpretation of GRB afterglows in terms of ambient interaction
models is complicated by several major uncertainties. For the jet 
model, the angular distribution of matter and energy inside the jet is 
not clear and the hydrodynamics of the jet-ambient medium interaction 
remains uncertain. Numerical simulations are beginning to address 
these issues  \cite{Ge01}. An open issue common to both the 
jet and wind models is the shape of the energy distribution of the 
radiative electrons. Fitting the afterglow data of several GRBs 
requires a curvature in the energy distribution. How the required 
curvature comes about is not understood. This problem is particularly 
severe for the cases that demand a flat distribution with $p < 2$. In 
such cases, the 
afterglow emission depends sensitively on the way in which the energy
distribution steepens at high energies, and the steepening has not been 
treated self-consistently. 
Another open issue is how the postshock energy is distributed among
electrons, protons, and magnetic field.
In addition, the spherical wind model faces 
an interesting dilemma: the wind model implies a massive star progenitor, 
and a collimated flow is expected if it must escape from the center of 
a star. To some extent, the spherical model can be justified by the 
slow apparent evolution of a jet in a wind  \cite{KP00a,Gou01}. 
It may also be possible that, upon passing through 
the star, the jet becomes uncollimated due to a sudden weakening of the 
lateral confinement. 
%% deconfinement of relativistic MHD jets and acceleration...AGNs, Crab

We conclude that there is evidence for two types of GRB afterglows in
different environments: a constant-density interstellar medium and 
a wind of possibly Wolf-Rayet star origin. The types are not immediately
distinguishable partly because, at an age of a few days, the preshock
wind density is comparable to an interstellar density. At an age of 
seconds to minutes, the preshock density is much higher for the wind
case, which could make the types more distinguishable. Pair production 
in the ambient medium could substantially modify the afterglow 
emission. Its effects are expected to be far greater in the wind case 
than in the ISM case  \cite{TM00,MRR01,B01}. 
In addition, the large difference in density for the 
two cases at small distances from the center of explosion affects the
dynamics of blast wave energization, and thus the prompt emission 
\cite{SP99,CL00}. While some theoretical 
work has been done on the prompt and early afterglow emission, much 
more is needed. Hopefully, rapid follow-up observations of GRB afterglows, 
to be enabled by HETE-2 and SWIFT, would put tighter constraints on the 
ambient interaction models and the nature of GRB progenitors. 

This work was supported in part by NASA grant NAG5-8130.

%INDEX%%%%%%%%%%%%%%%%%%%%%%%%%%%%%%%%%%%%%%%%%%%%%%%%%%%%%%%%%%%%%%%
% Please check with the editor of your book whether he plans to
% include a "mutual" subject index - if so, please code your entries
% in the standard syntax. For your own purposes you may print your
% "personal" index by using the following commands:
%
%\clearpage
%\addcontentsline{toc}{section}{Index}
%\flushbottom
%\printindex
%%%%%%%%%%%%%%%%%%%%%%%%%%%%%%%%%%%%%%%%%%%%%%%%%%%%%%%%%%%%%%%%%%%%%


\begin{thebibliography}{8.}
\addcontentsline{toc}{section}{References}


%\bibitem[]{}
%Akerlof, C. W., et al. 1999, Nature, 398, 400

\bibitem{Ae00}
E. Amato, M. Salvati, R. Bandiera, F. Pacini, L. Woltjer: A\&A
\textbf{359}, 1107 (2000)

\bibitem{Ane00}
L.A. Antonelli, et al.: ApJ \textbf{545}, L39 (2000)

\bibitem{Bh01}
D. Bhattacharya: Bull. Astr. Soc. India  \textbf{29}, 107 (2001)

\bibitem{B01}
A.M. Beloborodov: ApJ, submitted (astro-ph/0103321) (2001)

\bibitem{Be00}
E. Berger, et al.: ApJ  \textbf{545}, 56 (2000) % 0301c

\bibitem{Be01a}
E. Berger, et al.: GCN No. 968 (2001)  %010222

\bibitem{Be01b}
E. Berger, et al.: ApJ  \textbf{556}, 556 (2001)  %0418

\bibitem{BM76}
R.D. Blandford, C.F. McKee:  Phys. Fluids \textbf{19}, 1130 (1976)

\bibitem{Be98}
J.S. Bloom,  et al.:  ApJ \textbf{508}, L21  (1998) %0703 


%\bibitem[]{}
%Bloom, J. S., et al. 1999a, \apj, 518, L1

\bibitem{Be99}
J.S. Bloom,  et al.:  Nature \textbf{401}, 453 (1999)


\bibitem{CTe99}
A.J. Castro-Tirado, et al.:  ApJ \textbf{511}, L85 (1999)

%\bibitem[]{}
%Castro-Tirado, A. J., et al. 1999, Science, 283, 2069


\bibitem{CTe01}
A.J. Castro-Tirado, et al.:  A\&A \textbf{370}, 398 (2001)  %1208

\bibitem{C98}
R.A. Chevalier:  ApJ \textbf{499}, 810 (1998) 

\bibitem{C01}
R.A. Chevalier: Second Rome GRB Workshop (astro-ph/0102212) (2001)

\bibitem{CF01}
R.A. Chevalier, C. Fransson:  ApJ \textbf{558}, L27 (2001) 

\bibitem{CI83} R.A. Chevalier, J.N. Imamura: ApJ
\textbf{270}, 554 (1983)

\bibitem{CL99}
R.A. Chevalier, Z.-Y. Li:  ApJ \textbf{520}, L29 (1999) 

\bibitem{CL00}
R.A. Chevalier, Z.-Y. Li:  ApJ \textbf{536}, 195 (2000) 

\bibitem{CD99}
J. Chiang, C.D. Dermer: ApJ  \textbf{512}, 699 (1999)

\bibitem{Ce97}
E. Costa, et al.:  Nature \textbf{387}, 783 (1997)

\bibitem{Ce99}
S. Covino, et al.: A\&A \textbf{348},  L1 (1999)

\bibitem{Ce01}
R. Cowsik, T.P. Prabhu, G.C. Anupama, B.C. Bhatt, D.K. Sahu, S. Ambika,
Padmakar, S.G. Bhargavi: Bull. Astr. Soc. India, in press 
(astro-ph/0104363) (2001)

\bibitem{DDR01}
S. Dado, A. Dar, A. De Rujula: preprint (astro-ph/0107367) (2001)

\bibitem{DC01}
Z.G. Dai, K.S. Cheng: ApJ \textbf{558}, L109 (2001)    

\bibitem{DL98}
Z.G. Dai, T. Lu:  MNRAS \textbf{298}, 87 (1998)

\bibitem{DL99}
 Z.G. Dai, T. Lu: ApJ  \textbf{519}, L155 (1999) 
 
\bibitem{DL01}
 Z.G. Dai, T. Lu: A\&A  \textbf{367}, 501 (2001)  %  0301C
 
\bibitem{DD00}
A. Dar, A. De Rujula: preprint (astro-ph/0008474) (2000)

 
\bibitem{EH98}
E. Ergma, E.P.J. van den Heuvel:
A\&A  \textbf{331}, L29 (1998)

\bibitem{EB00}
A.A. Esin, R.D. Blandford: ApJ  \textbf{534}, L151 (2000)

\bibitem{Fe97}
D.A. Frail, S.R. Kulkarni, L. Nicastro, M. Feroci, G.B. Taylor:
 Nature \textbf{389}, 261 (1997)

\bibitem{Fe98}
D.A. Frail, S.R. Kulkarni, D.S. Shepherd, E. Waxman:
 ApJ  \textbf{502}, L119 (1998)

 
%\bibitem[]{}
%Frail, D. A.,  Kulkarni, S. R., Bloom, J. S., Djorgovski, S. G.
% et al. 1998c, GCN Circ. 141

%%\bibitem[]{}
%Frail, D. A., Kulkarni, S. R., Sari, R., Taylor, G. B., Shepherd, D. S.,
%Bloom, J. S., Young C. H., Nicastro, L., \& Masetti, N. 1999a, ApJ, in press
%(astro-ph/9910060)

\bibitem{FWK00}
D.A. Frail, E. Waxman, S.R. Kulkarni:
 ApJ \textbf{537}, 191 (2000)
 
\bibitem{Fe00a}
D.A. Frail, et al.: ApJ  \textbf{534}, 559 (2000)  % 0519

\bibitem{Fe00b}
D.A. Frail, et al.: ApJ  \textbf{538}, L129 (2000)   %1216

\bibitem{Fre01}
D.A. Frail, et al.: ApJ, submitted (astro-ph/0108436) (2000)    

% \bibitem[]{}
%Frontera, F., et al. 1998, A\&A, 334, L69

% \bibitem[]{}
%Frontera, F., et al. 1999, ApJS, in press
%astro-ph/9911228

%\bibitem[]{}
%Fruchter, A. S., et al. 1999a, \apj, 516, 683

%\bibitem[]{}
%Fruchter, A. S., et al. 1999b, preprint (astro-ph/9903236)

\bibitem{Fre00}
A.S. Fruchter,  et al.:  ApJ \textbf{545}, 664 (2000)

%\bibitem[]{}
%Fruchter, A. S., Ferguson, H., Pepper, J., Gibbons, R., Sahu, K.,  Pian, E.,
% et al. 1999c, GCN Circ. 386

\bibitem{FWH99}
C.L. Fryer, S.E. Woosley, D.H. Hartmann:  ApJ  \textbf{526}, 152 (1999)

%\bibitem[]{}
%Fryer, C. L., Woosley, S. E., Herant, M., \& Davies, M. B.
% 1999b, \apj, 520, 650

\bibitem{Fe01}
J.U. Fynbo, et al.: A\&A  \textbf{373}, 796 (2001) % 0926

\bibitem{Ge98a}
T.J. Galama, et al.:  ApJ \textbf{497}, L13 (1998) % 0508 alpha?

\bibitem{Ge98b}
T.J. Galama, R.A.M.J. Wijers, M. Bremer, P.J. Groot, R.G. Strom, C.
Kouveliotou, J. van Paradijs:  ApJ \textbf{500}, L97 (1998)  % 0508 cooling freq?

%\bibitem[]{}
%Galama, T. J., et al. 1998c, \apj, 500, L101

%\bibitem[]{}
%Galama, T. J., et al. 1998d, Nature, 395,670

%\bibitem[]{}
%Galama, T. J., et al. 1998e, GCN Circ. 127

%%\bibitem[]{}
%Galama, T. J., et al. 1999a, Nature, 398, 394

%\bibitem[]{}
%Galama, T. J., et al. 1999b, ApJ, submitted (astro-ph/9907264)

\bibitem{Ge00a}
T.J. Galama, et al.: ApJ  \textbf{536}, 185 (2000)  %sn0228

\bibitem{Ge00b}
T.J. Galama, et al.: ApJ  \textbf{541}, L45 (2000)  %1208

\bibitem{GS96}
G. Garc\'ia-Segura, N. Langer, M.-M. MacLow:
A\&A \textbf{316}, 133 (1996)

%\bibitem[]{}
%Ghisellini, G. 1999, preprint (astro-ph/9907376)

\bibitem{GL99}
G. Ghisellini, D. Lazzati: MNRAS  \textbf{309}, L7 (1999)   

\bibitem{Gou01}
L.-J. Gou, Z.G. Dai, Y.F. Huang, T. Lu: A\&A  \textbf{368}, 464 (2001) % jetwind

\bibitem{GPS99}
J. Granot, T. Piran, R.  Sari:  ApJ  \textbf{527}, 236 (1999)

\bibitem{GPS00}
J. Granot, T. Piran, R.  Sari:  ApJ  \textbf{534}, L163 (2000)

\bibitem{GS01}
J. Granot, R. Sari: preprint (astro-ph/0107009) (2001)

\bibitem{Ge01}
J. Granot, M. Miller, T. Piran, W.-M. Suen, P.A. Hughes:
2nd Rome GRB Workshop (astro-ph/0103038) (2001)  %jet

%\bibitem[]{}
%Groot, P. J., et al. 1998, \apj, 502, L123


%\bibitem[]{}
%Halpern, J. P., Kemp, J., Piran, T., \& Bershady, M. A. 1999,
%\apj, 517, L105

\bibitem{He00}
J.P. Halpern, et al.: ApJ \textbf{543}, 697 (2000) % 1216

\bibitem{He99}
F.A. Harrison et al.:  ApJ \textbf{523}, L121  (1999)

\bibitem{He01}
F.A. Harrison et al.:  ApJ \textbf{559}, 123  (2001)


\bibitem{Ho00}
S. Holland, G. Bj\"ornsson, J. Hjorth, B. Thomsen: A\&A  \textbf{364}, 467 (2000)

\bibitem{Ho01}
S. Holland, et al.: A\&A  \textbf{371}, 371 (2001)   0703

%\bibitem[]{}
%in 't Zand, J. J. M., et al. 1998, \apj, 505, L119

%\bibitem[]{}
%in 't Zand, J. J. M., Heise, J., van Paradijs, J., \& Fenimore, E. E. 1999,
%\apj, 516, L57

\bibitem{ie01}
J.J.M. in 't Zand,  et al.:  ApJ, in press (astro-ph/0104362)  (2001)

%\bibitem[]{}
%Israel, G. L., et al. 1999, A\&A, 348, L5

\bibitem{Je01}
A.O. Jaunsen,  et al.: ApJ \textbf{546}, 127 (2001)

\bibitem{K94}
J.I. Katz:  ApJ, \textbf{422}, 248 (1994)

\bibitem{Ke00}
S. Klose, et al.:  ApJ, \textbf{545}, 271 (2000)  %0418

%\bibitem[]{}
%Kulkarni, S. R. 1999, talk at ITP Conf. on Gamma-Ray Bursts and their Afterglows
%(http://www.itp.ucsb.edu/online/gamma\_c99/kulkarni/)

%\bibitem[]{}
%Kulkarni, S. R., Frail, D. A., Wieringa, M. H., Ekers, R. D.,
%Sadler, E. M., Wark, R. M., Higdon, J. L., Phinney, E. S., \& Bloom, J. S.
%1998, Nature, 395, 663

\bibitem{Ke98}
S.R. Kulkarni,  et al.:  Nature \textbf{395}, 663 (1998)

\bibitem{Ke99}
S.R. Kulkarni,  et al.:  Nature \textbf{398}, 389 (1999)


\bibitem{KP00a}
P. Kumar, A. Panaitescu:  ApJ \textbf{541}, L9 (2000) %jet trans?

\bibitem{KP00b}
P. Kumar, A. Panaitescu:  ApJ \textbf{541}, L51 (2000) %naked grb

\bibitem{Ku00}
E. Kuulkers, et al.: ApJ \textbf{538}, 638 (2000)

\bibitem{LC99}
Z.-Y. Li, R.A. Chevalier:  ApJ \textbf{526}, 716 (1999)

\bibitem{LC01}
Z.-Y. Li, R.A. Chevalier:  ApJ \textbf{551}, 940  (2001)

\bibitem{LW00}
M. Livio, E. Waxman: ApJ \textbf{538}, 187 (2000)

%\bibitem[]{}
%MacFadyen, A., \& Woosley, S. E. 1999, \apj, 524, 262

\bibitem{MWH01}
A.I. MacFadyen, S.E. Woosley, A. Heger: ApJ  \textbf{550}, 410 
 (2001)

\bibitem{Me01}
N. Masetti, et al.:  A\&A  \textbf{374}, 382 (2001)

\bibitem{MRR01}
P. M\'esz\'aros, E. Ramirez-Ruiz, M.J. Rees: ApJ \textbf{554}, 660 (2001)

\bibitem{MR97}
P. M\'esz\'aros, M.J. Rees:   ApJ \textbf{476}, 232 (1997)

\bibitem{MR99}
P. M\'esz\'aros, M.J. Rees:   MNRAS \textbf{306}, L39 (1999) %edge of jet?

\bibitem{MR01}
P. M\'esz\'aros, M.J. Rees: ApJ \textbf{556}, L37 (2001) %collapsar bubble 

\bibitem{MRW98}
P. M\'esz\'aros, M.J. Rees, R.A.M.J. Wijers:   ApJ \textbf{499}, 301 (1998)

\bibitem{Me97}
M.R. Metzger, et al.: Nature  \textbf{387}, 878 (1997)

\bibitem{MSB00}
R. Moderski, M. Sikora, T. Bulik:  ApJ \textbf{529}, 151 (2000)

\bibitem{P98}
B. Paczy\'nski:  ApJ \textbf{494}, L45 (1998)

\bibitem{P01}
A. Panaitescu: ApJ \textbf{556}, 1002 (2001)

\bibitem{PK00}
A. Panaitescu, P. Kumar: ApJ  \textbf{543}, 66 (2000)  

\bibitem{PK01a}
A. Panaitescu, P. Kumar: ApJ  \textbf{554}, 667 (2001)  

\bibitem{PK01b}
A. Panaitescu, P. Kumar: ApJ \textbf{560}, L49  (2001) % 1208?

\bibitem{PK01c}
A. Panaitescu, P. Kumar: ApJ, submitted (astro-ph/0109124)  (2001) % 1208?

\bibitem{PM98}
A. Panaitescu, P. M\'esz\'aros:  ApJ \textbf{501}, 772 (1998)  %IC?

\bibitem{PMR98}
A. Panaitescu, P. M\'esz\'aros, M.J. Rees:  ApJ \textbf{503}, 315 (1998)

%\bibitem[]{}
%Panaitescu, A., \& M\'esz\'aros, P. 1999, \apj, 526, 707

\bibitem{Pie01}
E. Pian, et al.: A\&A  \textbf{372}, 456 (2001)


%\bibitem[]{}
%Piran, T. 1999, Phys. Rep., 314, 575

\bibitem{PG01}
T. Piran, J. Granot:  Second Rome GRB meeting (astro-ph/0107009) (2001)

\bibitem{Pe98}
L. Piro,  et al.:  A\&A \textbf{331}, L41 (1998)  %0508 XR?

\bibitem{Pe99}
L. Piro,  et al.:  ApJ \textbf{514}, L73 (1999)  %0508 XR line

\bibitem{Pire00}
L. Piro, et al.: Science  \textbf{290}, 955 (2000)  %1216 line

\bibitem{Pire01}
L. Piro, et al.: ApJ  \textbf{558}, 442 (2001)  %0926

%\bibitem[]{}
%Piro, L., et al. 1999, \apj, 514, L73

\bibitem{Pre01}
P.A. Price, et al.: ApJ  \textbf{549},  L7  (2001)

\bibitem{RR00}
E. Ramirez-Ruiz, L.M. Dray, P. Madau, C.A. Tout: MNRAS,  submitted 
(astro-ph/0012396) (2000)

%\bibitem[]{}
%Rees, M. J., \& M\'esz\'aros, P. 1998, ApJ, 496, L1

\bibitem{Rei99}
D.E. Reichart:  ApJ  \textbf{521}, L111 (1999)

\bibitem{Rei01}
D.E. Reichart:  ApJ  \textbf{554}, 643 (2001)

\bibitem{Rei01b}
D.E. Reichart:  ApJ  submitted (astro-ph/0107546) (2001)

\bibitem{R97}
J.E. Rhoads: ApJ \textbf{487}, L1 (1997)

\bibitem{R99}
J.E. Rhoads: ApJ \textbf{525}, 737 (1999)

\bibitem{R01}
J.E. Rhoads: Ninth Marcel Grossman Meeting (astro-ph/0103028)  (2001)

\bibitem{RF01}
J.E. Rhoads, A.S. Fruchter: ApJ  \textbf{546}, 117 (2001)


\bibitem{Se01a}
R. Sagar, S.B. Pandey, V. Mohan, D. Bhattacharya, A.J. Castro-Tirado:
 Bull. Astr. Soc. India \textbf{29}, 1 (2001)  0926

\bibitem{Se01b}
R. Sagar, et al.: Bull. Astr. Soc. India \textbf{29}, 91 (2001)  0222

%\bibitem[]{}
%Sahu, K. C., et al. 1997, Nature, 387, 476

%\bibitem[]{}
%Sari, R. 1997, \apj, 489, L37

%\bibitem[]{}
%Sari, R. 1998, \apj, 494, L49

\bibitem{S99}
R. Sari: ApJ  \textbf{524}, L43 (1999)

\bibitem{SE01}
R. Sari,  A.A. Esin:
ApJ  \textbf{548}, 787 (2001)

%\bibitem[]{}
%Sari, R. \& Piran, T. 1995, ApJ, 455, L143

\bibitem{SP99}
R. Sari, T. Piran:  ApJ \textbf{517}, L109 (1999)

%\bibitem[]{}
%Sari, R. \& Piran, T. 1999b, ApJ, 520, 641

\bibitem{SPH99}
R. Sari, T. Piran, J.P. Halpern: 
ApJ  \textbf{519}, L17 (1999)

\bibitem{SPN98}
R. Sari, T. Piran, R. Narayan:  
ApJ  \textbf{497}, L17 (1998)

%\bibitem[]{}
%Sedov, L. I. 1959, Similarity and Dimensional Methods in Mechanics
%(New York: Academic)

%\bibitem[]{}
%Smith, I. A., et al. 1999, A\&A, 347, 92

\bibitem{So01}
P.M. Solomon: in Starburst Galaxies: Near and Far (astro-ph/0101482) (2001)

\bibitem{S01}
V.V. Sokolov: Second Rome GRB Workshop (astro-ph/0102492) (2001)

\bibitem{Se98}
V.V. Sokolov, A.I. Kopylov, S.V. Zharikov, M. Feroci, 
L. Nicastro, E. Palazzi:  A\&A \textbf{334}, 117 (1998)

\bibitem{Se99}
K.Z. Stanek, P.M. Garnavich, J. Kaluzny, W. Pych, I. Thompson:
ApJ  \textbf{522}, L39 (1999)

\bibitem{Se01}
K.Z. Stanek, et al.:
ApJ,  in press (astro-ph/0104329) (2001)

%\bibitem[]{}
%Taylor, G. B., Frail, D. A., Kulkarni, S. R., Shepherd, D. S., Feroci, M.,
%\& Frontera, F. 1998, \apj, 502, L115

\bibitem{TM00}
C. Thompson, P. Madau: ApJ \textbf{538}, 105 (2000)

\bibitem{vPe97}
J. van Paradijs, et al.: Nature \textbf{386}, 686 (1997)

\bibitem{Ve01}
M. Vietri, G.Ghisellini, D. Lazzati, F. Fiore, L. Stella: ApJ \textbf{550}, L43 (2001)

\bibitem{Ve99}
P.M. Vreeswijk, et al.: ApJ  \textbf{523},  171 (1999) % 0703

\bibitem{W97a}
E. Waxman: ApJ \textbf{485}, L5  (1997)

\bibitem{W97b}
E. Waxman: ApJ \textbf{489}, L33 (1997)

%\bibitem{W97c}
%E. Waxman:   ApJ \textbf{491}, L19 (1997)

%\bibitem[]{}
%Waxman, E., Kulkarni, S. R., \& Frail, D. A. 1998, ApJ, \textbf{497, 288

\bibitem{WL99}
E. Waxman, A. Loeb: ApJ \textbf{515}, 721 (1999)

\bibitem{WL98}
D.M. Wei, T. Lu: ApJ \textbf{505}, 252 (1998)   

\bibitem{WL00}
D.M. Wei, T. Lu: ApJ \textbf{541}, 203 (2000)  
 
\bibitem{WL01}
D.M. Wei, T. Lu: preprint (astro-ph/0107371) (2001)   

\bibitem{WPM01}
K. Weiler, N. Panagia, M. Montes: ApJ, in press (astro-ph/0106131) (2001)   

\bibitem{We00}
C. Weth, P. M\'esz\'aros, T. Kallman, M.J. Rees: ApJ \textbf{534}, 581 (2000)  
 
\bibitem{WG99}
R.A.M.J. Wijers, T.J. Galama:  ApJ \textbf{523}, 177 (1999)

\bibitem{WRM97}
R.A.M.J. Wijers, M.J. Rees, P. M\'esz\'aros:  MNRAS \textbf{288}, L51 (1997)

\bibitem{We99}
R.A.M.J. Wijers, et al.:  ApJ \textbf{523}, L33 (1999)

\bibitem{W93}
S.E. Woosley:  ApJ \textbf{405}, 273 (1993)


%\bibitem[]{}
%Woosley, S. E., Eastman, R. G.,  \& Schmidt, B. P. 1999, ApJ, 

\bibitem{Ye99}
A. Yoshida, et al.: A\&AS \textbf{138}, 433  (1999)
%516, 788

\bibitem{Ye01}
S.A. Yost, et al.: Second Rome GRB Workshop (astro-ph/0107556) (2001)

%


\end{thebibliography}
\end{document}